# From Voice to Leverage in Data Governance

## *Building Enforceable Social License for Data Re-Use*

Stefaan G. Verhulst, Adam Zable, Begoña González Otero, and Jacqueline Lu
*The GovLab, New York University · University of Toronto, School of Cities*

## Abstract

Contemporary data governance has converged on participation. Citizens' assemblies, deliberative polls, co-design workshops, and community advisory panels are now routinely suggested to build trust and secure a *social license* for the re-use of data. Yet this consensus rests on an unexamined assumption: that giving affected communities a voice will, on its own, change what data holders do. This paper argues that the field needs to complement voice with *leverage*—the capacity of individuals and communities to make their participation consequential by attaching credible legal, institutional, contractual, economic, or reputational costs to being ignored. Drawing on labor-relations bargaining theory, the sociology of disputes, and James C. Scott's account of legibility, we develop leverage as an analytic category defined by three properties: it is counterfactual, enforceable, and durable. We further argue that leverage presupposes *epistemic legibility*—affected publics can rarely impose costs on a system they cannot see, document, or contest—and that most existing transparency instruments equip regulators and procurers rather than the governed. We then map six sources from which communities and data subjects can derive leverage (legal and regulatory; institutional decision; contractual; economic; data access; and political and reputational), examine the intermediary structures that aggregate and sustain it, and propose a six-dimension evaluative framework—obligation, enforceability, consequence, monitoring equipment, revocability, and institutional durability—for auditing whether any social license process actually redistributes power. The paper's central claim is that a social license worth the name must share the properties of a legal license: conditions, a licensor, and enforceable remedies for breach.

## Table of Contents

## I. Introduction

Over the past decade, both the practice and the scholarship of data governance have introduced a family of concepts, such as *social license*[1], *data stewardship*[2], *data commons*[3], and *digital self-determination*[4], among others, as alternatives to purely legal or market-based approaches to governing how data is re-used. Across this literature runs a common conviction: legitimacy depends not only on whether data *can* be used, but on whether those affected have a meaningful opportunity to shape *how* it is used. This is a genuine advance over the twin defaults of the previous era, notice-and-consent on one side and unfettered proprietary control on the other, each of which had come to look increasingly hollow as data moved from a by-product of transactions to a primary input for artificial intelligence.

Yet an important question remains largely unanswered. What happens *after* participation and deliberation? Under what conditions does community and public input actually change institutional behavior, rather than merely decorating a decision that has already been taken? The problem has two layers. The first concerns consequence: the mere existence of a participatory mechanism is not the same as its efficacy, and an invitation to deliberate can coexist comfortably with an institution's freedom to disregard whatever is said. The second concerns capacity: a mechanism may be formally available yet practically out of reach, so that participation reproduces the very asymmetry it was meant to correct. This second layer forces a prior question that the participation literature rarely asks: what must affected publics be able to *see and understand* in the first place, before meaningful participation is possible at all?

[1] Stefaan Verhulst and Adam Zable, Operationalizing a Social License for Data Re-Use: A Facilitator's Guide and Worksheet — Questions to Signal and Capture Community Preferences and Expectations (New York: The GovLab / Open Data Policy Lab, 2026), https://tools.opendatapolicylab.org/files/Data-Re-Use-Worksheet.pdf.

[2] Stefaan Verhulst, “Data Stewardship Decoded: Mapping Its Diverse Manifestations and Emerging Relevance at a Time of AI,” SSRN Scholarly Paper no. 5124555 (February 2025), https://papers.ssrn.com/sol3/papers.cfm?abstract_id=5124555.

[3] Stefaan Verhulst, Andrew J. Zahuranec, Hannah Chafetz, Leona Verdadero, and Jennifer Hansen, Data Commons: Frequently Asked Questions (New Commons Incubator, in collaboration with the Open Data Policy Lab at The GovLab, 2026), https://opendatapolicylab.org/articles/data-commons-faq-report/.

[4] International Network on Digital Self-Determination, accessed August 25, 2026, https://www.idsd.network/.

This paper argues that contemporary debates have focused disproportionately on *voice* while neglecting *leverage*.[5] Building on our earlier work on social license for data re-use,[6] we develop the concept of *social leverage*: the capacity of individuals and communities to make their participation consequential by creating legal, institutional, contractual, financial, or reputational incentives that make their preferences costly to ignore. Leverage, in this sense, is not only a matter of mechanisms of consequence but of *equipment*: communities can rarely impose costs on a system they cannot see, document, or track. Part of designing for leverage, we argue, is therefore building the practical instruments that render data systems legible and contestable to the people they affect.

Our argument proceeds in five further steps. Part II diagnoses the participation consensus and its blind spot, situating the social license discourse within a longer lineage of participation theory that reaches back to Arnstein's ladder. Part III defines leverage analytically, distinguishing it from voice and exit and grounding it in labor-relations bargaining theory; it also introduces the precondition of epistemic legibility. Part IV develops a taxonomy of six sources of leverage, illustrated throughout with concrete instruments and cases. Part V examines intermediary structures—data trusts, cooperatives, commons, unions, and coalitions—as the organizational vehicles through which leverage is aggregated and made durable. Part VI turns the analysis into a design and evaluation framework. Rather than treating engagement as an end in itself, we propose that effective data governance must deliberately connect participation with mechanisms of accountability and consequence, extending the emerging theory of social license from one centered on consultation and trust toward one grounded in shared power, institutional design, and durable governance.

## II. The Participation Consensus and Its Blind Spot

Few ideas enjoy greater support in contemporary data governance than participation. Whether the subject is health-data research, smart cities, the training data behind large AI models, or public-sector data sharing, the prescription is remarkably uniform: engage the affected, deliberate carefully, and ensure that all voices and interests are taken into account. An entire repertoire of methods has grown up around this ambition—citizens' assemblies and juries convened on questions of data, deliberative polling, public dialogues, co-design workshops, and standing engagement panels attached to data institutions. This turn draws energy from a broader revival of deliberative and participatory democracy, in which structured public reasoning is treated as a source of legitimacy distinct from both markets and representative voting.[7]

[5] Nicholas Vincent, Hanlin Li, Nicole Tilly, Stevie Chancellor, and Brent Hecht, “Data Leverage: A Framework for Empowering the Public in Its Relationship with Technology Companies,” in Proceedings of the 2021 ACM Conference on Fairness, Accountability, and Transparency (FAccT '21) (New York: ACM, 2021), 215–27, https://doi.org/10.1145/3442188.3445885; preprint at arXiv:2012.09995, https://arxiv.org/abs/2012.09995.

[6] Verhulst and Zable, Operationalizing a Social License for Data Re-Use; see also Open Data Policy Lab, “A Facilitator's Guide to Establishing a Social License for Data Reuse,” February 10, 2026, https://opendatapolicylab.org/articles/new-publication-a-facilitators-guide-to-establishing-a-social-license-for-data-reuse/.

The concept of *social license* has become the organizing frame for this work. Borrowed from the extractive industries, where it named the informal, ongoing acceptance a mining or forestry operation must secure from the communities in whose midst it operates,[8] it captures the intuition that lawful access to data is not the same as legitimate access: data re-use requires an ongoing, informal permission from the communities whose data—and whose lives—are implicated. The collapse of England's care.data programme in 2014, abandoned in the face of public opposition despite resting on a statutory basis,[9] remains the canonical demonstration that this license is both necessary and revocable.

Yet the social license discourse conceals a blind spot that it shares with participation theory more broadly. Both are overwhelmingly concerned with the *input* side of the process: whose voices are heard, how representative they are, whether trust is built, whether expectations are surfaced. Far less attention is paid to what happens after the deliberation ends. How does input become consequence? What compels the data holder or re-user—a technology company, a hospital system, a city government, a research consortium—to do anything with what it has heard? And what can a community actually do, short of a withdrawal of trust, when its stated conditions are simply ignored?

There is a second, quieter dimension to this blind spot: even where communities are consulted, they are seldom equipped. A landscape scan by Helpful Places[10] of more than fifty AI and data legibility instruments — disclosure registers, transparency standards, model cards, and the like — found that affected publics are almost always named as the *object* of transparency but rarely treated as its *audience*. Only about a fifth of the instruments surveyed delivered any usable, public-facing artefact to the people they described, and almost none gave those people a way to track whether a system changed after the moment of disclosure.[11] The tools built in the name of transparency, in other words, largely equip the choosers—regulators, procurers, deployers—rather than the affected.

---

[7]On the deliberative-democratic foundations of this turn, see John S. Dryzek, Deliberative Democracy and Beyond: Liberals, Critics, Contestations (Oxford: Oxford University Press, 2000); and, for a critical synthesis of the participatory-governance tradition, Archon Fung, "Varieties of Participation in Complex Governance," Public Administration Review 66, s1 (2006): 66–75, https://doi.org/10.1111/j.1540-6210.2006.00667.x.

[8]The concept originates in mining-sector debates of the 1990s; for its migration into information governance, see Jenifer Sunrise Winter and Elizabeth Davidson, "Governance of Artificial Intelligence and Personal Health Information," Digital Policy, Regulation and Governance 21, no. 3 (2019): 280–90, https://doi.org/10.1108/DPRG-08-2018-0048.

[9]Sigrid Sterckx, Vojin Rakic, Julian Cockbain, and Pascal Borry, "'You Hoped We Would Sleep Walk into Accepting the Collection of Our Data': Controversies Surrounding the UK care.data Scheme and Their Wider Relevance for Biomedical Research," Medicine, Health Care and Philosophy 19, no. 2 (2016): 177–90, https://doi.org/10.1007/s11019-015-9661-6.

[10] Jyoti Singh, Charles Finley, and Jacqueline Lu, "Named, But Not Addressed: Affected Publics in the AI Transparency Landscape" (Helpful Places, working paper, 2026). The scan was undertaken as part of an official Standards Council of Canada Workshop Agreement development process, in response to a gap identified in Canada's Data Governance and Standardization Roadmap: that existing AI transparency and disclosure instruments are seldom designed to be understood by the people they describe, leaving affected publics named as their subject but rarely equipped as their audience.

Without an answer to these questions, social license risks becoming what Sherry Arnstein diagnosed more than half a century ago as the middle rungs of her ladder of citizen participation: informing, consultation, and placation—*voice granted precisely because power is not*.[12] Worse, it risks functioning as a communications strategy—license as something an organization *earns* through the performance of engagement, rather than something a community *holds* and can enforce. The critical-data-studies literature has repeatedly warned that participatory language can be captured to legitimate rather than constrain, a dynamic sometimes described as *participation-washing*.[13] The missing concept, until now, is leverage.

## III. What Leverage Means

Leverage is the capacity to make one's input costly to ignore. It is helpful to locate it against two neighboring concepts inherited from Albert Hirschman. *Voice* expresses preferences; *effective voice*[14] produces a defined institutional effect; and *exit*[15] withdraws participation, cooperation, data, or legitimacy. Leverage is what makes voice or exit consequential, by attaching credible costs to being ignored. Where participation asks to be heard, leverage ensures that not listening carries a price—legal, financial, institutional, or reputational. Three features distinguish leverage from voice.

First, leverage is *counterfactual*: it changes what the data holder would otherwise do, rather than merely informing or legitimating a decision already within their discretion. A consultation that leaves the decision-maker free to proceed exactly as planned has produced voice but not leverage. Second, leverage is *enforceable*: it does not depend on the continued goodwill of the more powerful party, but operates through mechanisms that hold even against that party's preferences—a court order, a contractual clause, a veto, a withheld payment, a revoked permission. Third, leverage is *durable*: it survives changes in personnel, shifts in political attention, and the turning of funding cycles, whereas invited engagement typically lasts only as long as the invitation.

---

[11]This landscape finding is consistent with a broader critique of transparency artefacts as compliance instruments; see Margot E. Kaminski, “Understanding Transparency in Algorithmic Accountability,” in The Cambridge Handbook of the Law of Algorithms, ed. Woodrow Barfield (Cambridge: Cambridge University Press, 2020), 121–38. On model cards' original framing, see Margaret Mitchell et al., “Model Cards for Model Reporting,” in Proceedings of the Conference on Fairness, Accountability, and Transparency (FAT* '19) (New York: ACM, 2019), 220–29, https://doi.org/10.1145/3287560.3287596.

[12]Sherry R. Arnstein, “A Ladder of Citizen Participation,” Journal of the American Institute of Planners 35, no. 4 (1969): 216–24, https://doi.org/10.1080/01944366908977225.

[13]Mona Sloane, Emanuel Moss, Olaitan Awomolo, and Laura Forlano, “Participation Is Not a Design Fix for Machine Learning,” in Proceedings of the 2nd ACM Conference on Equity and Access in Algorithms, Mechanisms, and Optimization (EAAMO '22) (New York: ACM, 2022), art. 1, https://doi.org/10.1145/3551624.3555285; see also Abeba Birhane et al., “Power to the People? Opportunities and Challenges for Participatory AI,” in EAAMO '22, art. 6, https://doi.org/10.1145/3551624.3555290.

[14]Seth Frey and Nathan Schneider, “Effective Voice: Beyond Exit and Affect in Online Communities,” arXiv preprint arXiv:2009.12470, September 25, 2020, https://arxiv.org/abs/2009.12470.

[15]Peter Levine, “Exit, Voice and Loyalty,” Civic Theory and Practice (blog), Tufts University, January 27, 2022, https://sites.tufts.edu/civicstudies/2022/01/27/exit-voice-and-loyalty/, discussing Albert O. Hirschman, Exit, Voice, and Loyalty: Responses to Decline in Firms, Organizations, and States (Cambridge, MA: Harvard University Press, 1970).

The vocabulary of *license* itself points in this direction, if taken seriously. A license, in law, is a conditional and revocable permission: it has terms, a licensor, and breach carries enforceable remedies. A *social* license should share these properties. The question “does this data re-use enjoy social license?” should therefore always be accompanied by a second: “through what mechanism could that license be conditioned, monitored, and withdrawn; and by whom?” Deliberation without leverage is decorative; leverage without deliberation is blind. The design challenge for data governance is to couple with them.

Labor scholars spent the better part of a century on precisely this problem. Their central answer (that bargaining power is the ratio between an opponent's cost of disagreeing with you and their cost of agreeing) remains the most serviceable definition of leverage available.[16] Chamberlain's formulation is useful because it is relational and manipulable: a party gains leverage not by asserting its preferences more loudly but by raising the other side's cost of ignoring them or lowering its own cost of holding out. A second lesson from the same tradition matters just as much: collaborative participation delivers for the weaker party only where that party retains an independent, countervailing capacity to sanction. Empowered participatory governance, Fung and Wright argue, depends on a rough balance of power that participation alone does not create.[17] Data governance has imported the deliberative half of this tradition while mostly leaving the power half behind.

## Legibility as the Precondition of Leverage

Before turning to the sources of that power, a precondition beneath all of them must be named: leverage without legibility is inert. None of the mechanisms that follow can be brought to bear against a system that its affected public cannot see, understand, or document. Legibility is not itself leverage, but it is the condition under which every other source becomes exercisable.

Legibility runs in two directions, and both must hold. The first is *outward*: the system must be perceptible; that data is being collected here, by this actor, for this purpose. The second is *reflexive*, and more often neglected: those affected must be able to recognize what they are entitled to do about it. The sociology of disputes has long shown that most injuries fall out not at the stage of remedy but far earlier, at the stage of *naming*: an experience must be recognized as an injurious one before it can be *blamed* on a responsible actor, and blamed before it can be *claimed* as a grievance.[18] Because so much attrition happens before a claim is ever articulated, legibility that stops at disclosure changes little.

---

[16]Neil W. Chamberlain, Collective Bargaining (New York: McGraw-Hill, 1951), chap. 10; discussed in Paul R. Hays, review of Collective Bargaining, by Neil W. Chamberlain, Indiana Law Journal 27, no. 2 (Winter 1952): art. 10, https://www.repository.law.indiana.edu/ilj/vol27/iss2/10.

[17]Archon Fung and Erik Olin Wright, “Thinking about Empowered Participatory Governance,” in Deepening Democracy: Institutional Innovations in Empowered Participatory Governance, ed. Archon Fung and Erik Olin Wright (London: Verso, 2003), 3–42, https://www.hks.harvard.edu/publications/thinking-about-empowered-participatory-governance.

This is why legibility cannot be equated with disclosure. Disclosure regimes routinely satisfy a legal duty without altering any behavior, and an explanation that discharges a duty may do nothing to produce understanding in its recipient.[19] Transparency describes an institution's disclosed behavior; legibility describes how the community perceives and understands that behavior. The distinction is captured by borrowing, and inverting, James C. Scott's concept of legibility. Where Scott describes how states render populations legible to themselves in order to tax, conscript, and administer them,[20] public-facing legibility inverts the direction of the gaze: it asks whether governance systems are understandable, trackable, and contestable to those they govern.

As with the power tradition above, legibility is best borrowed rather than reinvented. Impact assessment, public participation, and privacy law have each built machinery to make institutional conduct visible, and that machinery can be inherited. What should not be inherited is the assumption that visibility suffices—the failure mode common to all three, and the one this paper has already diagnosed in participation. The design task is to render systems legible in a form that carries through to consequence. The open standard Digital Trust for Places and Routines (DTPR), which signals how data, sensors, and AI are used in shared physical space, is instructive precisely because it pairs legibility with accountability rather than offering the former in place of the latter.[21]

## IV. Six Sources of Leverage in Data Re-Use

If leverage is the capacity to impose consequences, the practical question becomes: from what sources can data subjects and affected communities derive it? Six families stand out, each defined by the underlying source of power or enforceability on which it rests—legal and regulatory authority; institutional decision rights; negotiated obligations; economic dependency; control over data or access; and political or reputational vulnerability. These sources may be exercised individually or in combination, and in practice they frequently overlap and reinforce one another. Each carries characteristic strengths and characteristic failure modes, which we set out below.

[18]William L. F. Felstiner, Richard L. Abel, and Austin Sarat, "The Emergence and Transformation of Disputes: Naming, Blaming, Claiming...," Law & Society Review 15, no. 3/4 (1980–81): 631–54, https://doi.org/10.2307/3053505.

[19]Omri Ben-Shahar and Carl E. Schneider, More Than You Wanted to Know: The Failure of Mandated Disclosure (Princeton, NJ: Princeton University Press, 2014); Lilian Edwards and Michael Veale, "Slave to the Algorithm? Why a 'Right to an Explanation' Is Probably Not the Remedy You Are Looking For," Duke Law & Technology Review 16 (2017): 18–84.

[20]James C. Scott, Seeing Like a State: How Certain Schemes to Improve the Human Condition Have Failed (New Haven, CT: Yale University Press, 1998).

[21]DTPR (Digital Trust for Places & Routines), stewarded by Helpful Places, accessed August 25, 2026, https://dtpr.io/.

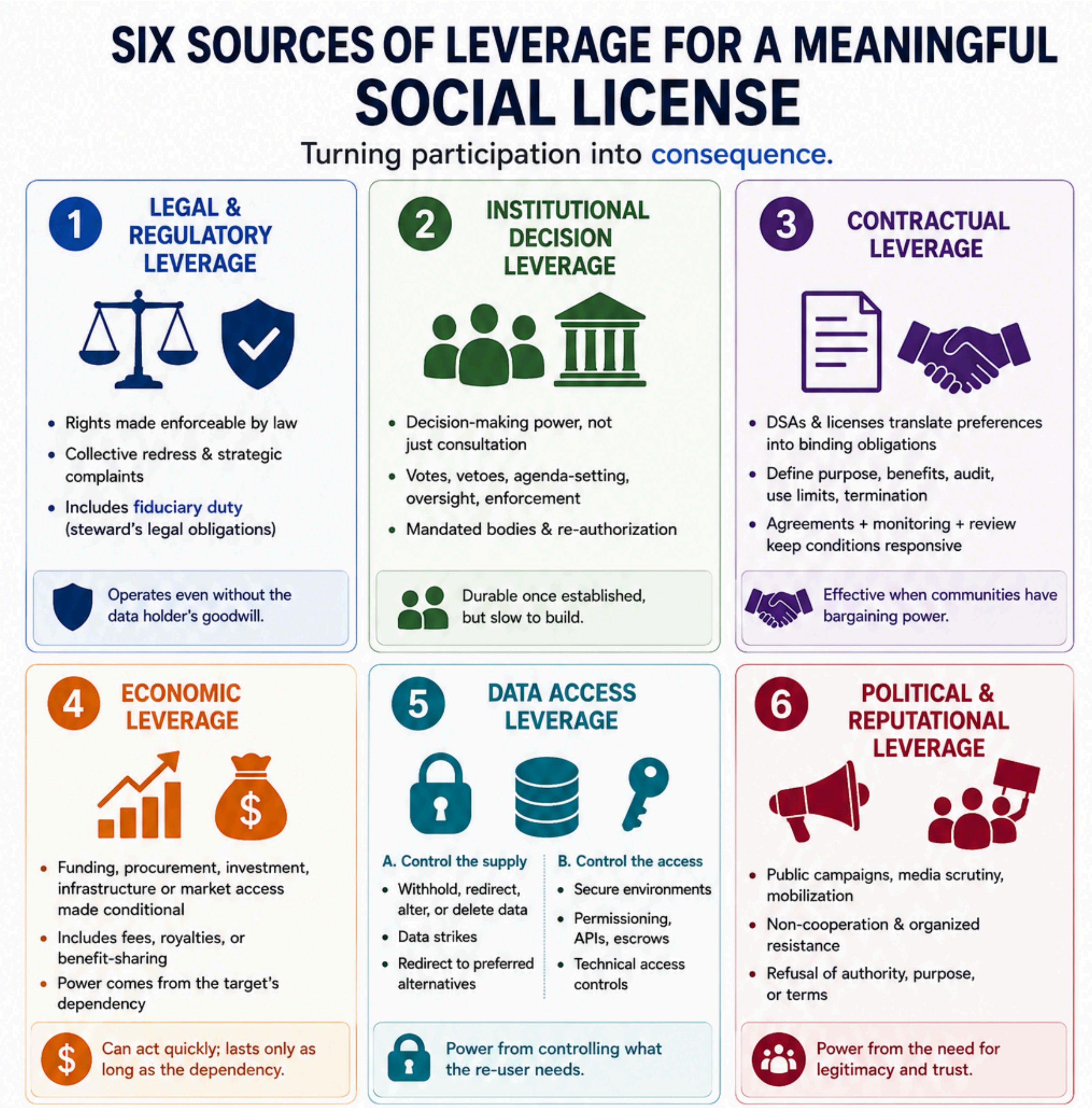

Figure 1: Six Sources of Leverage

## A. Legal and Regulatory Leverage

Data protection law already converts some interests into enforceable entitlements—consent, access, erasure, and objection chief among them. On their own, individual rights are a thin form of leverage; their force multiplies when aggregated. Collective redress, from the EU's Representative Actions Directive to strategic complaints brought by civil society organizations, adds this collective dimension, so that a single case or regulatory intervention can discipline an entire sector.[22] Indigenous data sovereignty movements have pushed the logic further still,[23] asserting collective rights over data as an extension of rights over territory and resources and codifying them in instruments such as the CARE Principles for Indigenous Data Governance.[24] Legal and regulatory leverage is valuable because it can operate independently of the data holder's goodwill. Its weaknesses are cost, delay, and an often individualistic architecture in which most data rights attach to persons[25] even though the harms and benefits of data re-use frequently accrue to groups.[26]

A distinct enforcement logic sits within this family: *fiduciary duty*. Where data is held in trust or under a stewardship mandate, duties of loyalty and care can bind the steward as a matter of law, enforceable by beneficiaries who never negotiated anything and carrying remedies—an account of profits, the removal of trustees—that ordinary contract does not offer. Proposals for "information fiduciaries" extend this logic to platforms that hold user data,[27] though the doctrine's reach is contested. Fiduciary leverage disciplines the steward where the other mechanisms in this family discipline the re-user.

### B. Institutional Decision Leverage

The second source is formal authority embedded within the governance architecture of data institutions. The crucial distinction is between advisory presence and decision-making power—between a public panel that is merely consulted and community representatives who hold voting rights, vetoes, agenda-setting authority, appointment or removal rights, or powers of investigation and remedy. Statutory consultation duties whose breach invalidates a decision, oversight bodies armed with enforcement powers, mandatory re-authorization requirements, and Māori and First Nations data governance bodies embedded within national statistical systems all belong to this family. Data stewards, too, can hold such leverage where they combine a defined mandate, meaningful authority, and genuine accountability to affected groups. Institutional decision leverage is durable and comparatively inexpensive to exercise once established, but it is slow to build and vulnerable to co-optation: a seat at the table can quietly become a reason to stop applying pressure outside it.

---

[22] Directive (EU) 2020/1828 of the European Parliament and of the Council of 25 November 2020 on representative actions for the protection of the collective interests of consumers and repealing Directive 2009/22/EC, OJ L 409, 4.12.2020, 1. The Directive allows collective actions to be brought against businesses if they breach EU law in a broad range of areas, such as data protection, travel and tourism, financial services, energy and telecommunication, with the aim of making collective redress more uniformly available to consumers across the EU. See further information at BEUC's website: https://www.beuc.eu/general/collective-redress.

[23] The GovLab, "Updated and Expanded Selected Readings on Indigenous Data Sovereignty," The GovLab Blog, October 12, 2020, updated October 11, 2021, https://blog.thegovlab.org/post/selected-readings-on-indigenous-data-sovereignty.

[24] Global Indigenous Data Alliance, "CARE Principles for Indigenous Data Governance" (2018), https://www.gida-global.org/careprinciples; see also Stephanie Russo Carroll et al., "The CARE Principles for Indigenous Data Governance," Data Science Journal 19, no. 1 (2020): art. 43, https://doi.org/10.5334/dsj-2020-043.

[25] Centre for Information Policy Leadership, The Limitations of Consent as a Legal Basis for Data Processing in the Digital Society (CIPL, December 2024), https://www.informationpolicycentre.com/wp-content/uploads/2024/12/cipl_bkl_limitations_of_consent_legal_basis_data_processing_dec24-4.pdf.

[26] Salomé Viljoen, "A Relational Theory of Data Governance," Yale Law Journal 131, no. 2 (2021): 573–654, https://papers.ssrn.com/sol3/papers.cfm?abstract_id=3727562.

[27] Jack M. Balkin, "Information Fiduciaries and the First Amendment," UC Davis Law Review 49, no. 4 (2016): 1183–1234; for a critique, see Lina M. Khan and David E. Pozen, "A Skeptical View of Information Fiduciaries," Harvard Law Review 133, no. 2 (2019): 497–541.

### C. Contractual Leverage

The third source is private ordering, which often provides the most immediately available route. Data sharing agreements (DSAs) and licensing agreements can translate community preferences into precise obligations governing purpose, benefit-sharing, audit, attribution, downstream use, and termination. An instructive example is the Nwulite Obodo Open Data License (NOODL), whose terms require users in high-income countries to provide a benefit or transfer value specified by the dataset's provider before using the data, while extending more permissive terms to users in developing countries.[28] Contract generally depends on prior bargaining power: the community must control something the re-user needs, and its willingness to walk away must be credible. Once established, however, an agreement can preserve that power by furnishing a shared reference point for accountability,[29] defining monitoring and dispute-resolution procedures, and pairing contractual remedies with technical safeguards and participatory oversight. Standardized clauses can further reduce transaction costs and improve legal clarity, while periodic review, amendment, suspension, and termination allow community-defined conditions to remain responsive over time.

### D. Economic Leverage

The fourth source is economic. Research funders, development banks, investors, and public buyers can make grants, disbursement, procurement eligibility, accreditation, infrastructure access, or market participation contingent on community authorization, benefit-sharing, or compliance with governance conditions. Its force derives from the target's dependence on financial support or economic opportunity. Public procurement is an especially underused instrument here: as the largest single purchaser in most economies, the state can embed governance conditions into the contracts on which vendors depend.[30] Economic arrangements often overlap with contractual and licensing leverage, as when an agreement conditions access or continued use on fees, royalties, or another form of benefit-sharing controlled by the affected community. Economic leverage can act quickly, but it lasts only as long as the dependency remains, and it concentrates in the hands of those who already hold capital, contracts, or market access.

[28]Alek Tarkowski, NOODL: An Experiment in Equitable Data Licensing — Promise and Limits (Open Future, April 21, 2026), https://openfuture.eu/publication/noodl-an-experiment-in-equitable-data-licensing-promise-and-limits/.

[29]Peter Addo, Adam Zable, Andrew J. Zahuranec, and Stefaan Verhulst, Reimagining Data Governance for AI, Technical Reports no. 78 (Paris: Agence Française de Développement, April 2025), https://www.afd.fr/en/ressources/reimagining-data-governance-ai.

[30]On procurement as a governance lever for algorithmic systems, see Deirdre K. Mulligan and Kenneth A. Bamberger, “Procurement as Policy: Administrative Process for Machine Learning,” Berkeley Technology Law Journal 34, no. 3 (2019): 773–852, https://doi.org/10.15779/Z38XG9FB0S.

### E. Data Access Leverage

The fifth source is control over data or the environments through which it is accessed, and it takes two forms. The first arises from direct control over the *supply* of data. Digital systems frequently depend on continuing contributions from users and communities,[31] creating leverage wherever those contributions can be withheld, redirected, or altered. Data strikes exploit this dependency by halting the flow of new data or withdrawing existing contributions—deleting data, blocking the collection of behavioral signals, refusing to supply ratings or reviews, or collectively logging off a platform.[32][33] A related strategy is to redirect valuable data toward preferred competitors or public-interest alternatives.[34] When participants stop using or paying for a service, these tactics shade into the economic effects of a boycott. The second form arises from *technical* control over access—through secure processing environments, permissioning systems, APIs, and data escrows[35]—each of which derives force from control over something the re-user needs in order to obtain or continue using the data.

### F. Political and Reputational Leverage

The sixth source arises from an institution's dependence on legitimacy, public support, reputation, and continued social or political cooperation. It may be exercised through public campaigns, organized non-cooperation, political mobilization, challenges to institutional authority, and other forms of resistance.[36] The political practice of *refusal* extends beyond withholding data: affected communities may also reject the authority of the collector, the legitimacy of the purpose, or the very terms on which participation is offered.[37]

[31]Vincent et al., "Data Leverage," 215–27.

[32]Nicholas Vincent, Brent Hecht, and Shilad Sen, "'Data Strikes': Evaluating the Effectiveness of a New Form of Collective Action against Technology Companies," in The World Wide Web Conference (WWW '19) (New York: ACM, 2019), https://doi.org/10.1145/3308558.3313742.

[33]"'Data Strike' Calls for Logging Off Social Media and Streaming Services on May 1st," Cybernews, April 22, 2026, https://cybernews.com/tech/data-strike-may/.

[34]Nicholas Vincent and Brent Hecht, "Can 'Conscious Data Contribution' Help Users to Exert 'Data Leverage' against Technology Companies?," Proceedings of the ACM on Human-Computer Interaction 5, no. CSCW1 (April 2021): art. 103, https://doi.org/10.1145/3449177.

[35]RadicalxChange Foundation, "Data Escrows," accessed August 25, 2026, https://www.radicalxchange.org/projects/data-escrows/.

[36]Renée Sieber and Roberta Du, "Citizen Resistance to AI" (presentation, Participatory AI Research & Practice Symposium (PAIRS) 2026, New Delhi and online, May 2026), https://aifortherestofus.ca/wp-content/uploads/2026/05/PAIRS-Sieber-2026.pdf.

### G. Leverage in Combination: The care.data Case

These sources rarely operate in isolation. The care.data controversy illustrates how several can combine into a single, decisive pressure.[38] Public criticism, media scrutiny, and declining trust in the NHS generated political and reputational pressure; patient opt-outs and general practitioners' threats to withhold records placed access to the data itself at risk; resistance from the British Medical Association and individual practitioners supplied institutional force; and disputes over confidentiality, consent, and statutory authority brought legal and regulatory leverage to bear. The combined pressure forced delays and substantive changes to the scheme, demonstrating how disparate sources of leverage can be knitted together into real institutional change. The lesson is that the sources are complementary and thus, the art of designing for leverage lies in their deliberate combination.

## V. Intermediaries: Aggregating and Sustaining Leverage

Cutting across these six sources are *intermediary structures*[39]—data trusts, data commons,[40] data cooperatives, data unions,[41] and data coalitions[42]—that aggregate, institutionalize, and deploy leverage. These are best understood not as a seventh source but as organizational vehicles. A data cooperative does not itself compel a data holder to do anything; it creates *associational* power by pooling rights and contributions, reducing coordination costs, representing members, monitoring compliance, financing enforcement, and making collective commitments credible. It may then negotiate licenses, pursue litigation or regulatory complaints, exercise governance rights, administer technical access controls, allocate benefits, or coordinate withdrawal.

The labor analogy makes the point precisely: a union is not the same thing as a collective bargaining agreement; it is the standing organization that makes negotiation, enforcement, and escalation possible over time. Intermediaries thus help solve the durability problem that one-off deliberative processes leave behind. Their standing risk is the mirror image of their promise: an intermediary without independent accountability to its community merely relocates the leverage problem to the question of who governs the intermediary—a concern that the data-trust literature has flagged as the central design challenge for any body that holds rights on others' behalf.[43]

[37] Jonathan Zong and J. Nathan Matias, "Data Refusal from Below: A Framework for Understanding, Evaluating, and Envisioning Refusal as Design," ACM Journal on Responsible Computing 1, no. 1 (2024): art. 10, https://doi.org/10.1145/3630107. On refusal as a sovereign practice more broadly, see Audra Simpson, Mohawk Interruptus: Political Life across the Borders of Settler States (Durham, NC: Duke University Press, 2014).

[38] Sterckx et al., "'You Hoped We Would Sleep Walk,'" 182–84.

[39] Stefaan Verhulst, "Orchestrating and Designing Data Collaboratives: What Governance Model Is Fit for Purpose?," SSRN Scholarly Paper no. 6364718 (March 7, 2026), https://papers.ssrn.com/sol3/papers.cfm?abstract_id=6364718.

[40] New Commons Incubator for Indigenous Languages and Cultures, accessed August 25, 2026, https://newcommons.ai/.

[41] RadicalxChange Foundation, "The Data Freedom Act," accessed August 25, 2026, https://www.radicalxchange.org/projects/the-data-freedom-act/.

[42] Jack Henderson and Matt Prewitt, Data Coalitions & Escrow Agents (RadicalxChange Foundation, June 2023), https://www.radicalxchange.org/updates/documents/data-coalitions-and-escrow-agents.pdf.

Intermediaries also depend on something prior to organization: a reliable, shared account of what a system or dataset actually does. A trust, cooperative, or coalition can only monitor compliance, verify a deployer's claims, or judge when practice has drifted far enough from what was promised to warrant escalation if it can draw on a stable, comparable record of the system in question. This is where a public disclosure standard such as DTPR does its work—not as leverage in itself, since it is neither counterfactual nor enforceable, but as the evidentiary layer on which associational power operates: the shared baseline that converts "we object" into "here is the commitment, here is the clause, and here is the deviation from it." Two features make such disclosure usable in practice. It is rarely a single document but a whole system of parts—a normative core, a way to assess against it, a public register, an oversight function, and a community of practice. And it is achievable rather than aspirational: in a 2026 municipal pilot, roughly nine in ten surveyed residents reported that the use of DTPR helped them understand the purpose of the AI system operating in a shared public space.[44]

How far each of these sources can be pushed—and how they interact within a given intermediary—remains a matter for further empirical research. The important point is that the menu exists, and that meaningful social license depends on deliberately connecting participation to one or more of these sources of consequence.

## VI. Designing for Leverage: An Evaluation Framework

Seen through this lens, the evaluation of any social license process should extend beyond inclusiveness, deliberative quality, and measured trust to a further set of questions. We propose six evaluative dimensions, each phrased as a question a designer, funder, or affected community can ask of any engagement process to determine whether it redistributes power or merely performs consultation.

Obligation. Does the engagement process generate formal, binding commitments, or does it end in a report that the data holder is free to shelve? Voice without consequence leaves the decision wholly within the holder's discretion.

Enforceability. Who holds standing to enforce those commitments, before which forum, and at what cost? Formal enforceability means little if the remedy is inaccessible in practice. A right that only well-resourced litigants can invoke is, for most affected publics, not effective remedy at all.

---

[43]Sylvie Delacroix and Neil D. Lawrence, "Bottom-Up Data Trusts: Disturbing the 'One Size Fits All' Approach to Data Governance," International Data Privacy Law 9, no. 4 (2019): 236–52, https://doi.org/10.1093/idpl/ipz014.

[44] Helpful Places. "Canada's AI strategy focuses on trust and transparency. But for whom?" June 30, 2026. https://www.helpfulplaces.com/insights/canada-s-ai-strategy-names-trust-and-transparency

Consequence. What concrete penalty, sanction, or other adverse consequence follows—and to whom—if the community's conditions are breached? Leverage requires that non-compliance carry an actual and attributable cost.

Monitoring capacity. Do affected groups have the instruments to see, verify, and monitor the system in the first place—and do those instruments remain in their hands after the process ends? This is the legibility precondition, restated as a design requirement.

Revocability. Can the license actually be conditioned, suspended, or revoked short of scandal or catastrophic collapse? A permission that can be withdrawn only by destroying the relationship is not a governance mechanism but a doomsday device.

Institutional durability. Does the process leave behind any durable structure, right, or resource—a seat, a clause, a fund, an intermediary capable of holding and exercising leverage over time—or does it evaporate when the deliberation ends?

The uncomfortable implication of this framework is that leverage is rarely given; it is conceded under pressure or built deliberately over time. Those who design social license processes for data re-use therefore face a choice. They can continue to perfect the machinery of voice—better sampling, better facilitation, better measurement of public attitudes—while leaving the translation into consequence to chance and goodwill. Or they can treat leverage as a design requirement of equal rank: pairing every engagement with an enforceable commitment, every panel with a vote, every condition with a remedy, every license with a credible means of revocation, and every claim of transparency with an instrument the affected public can actually use to hold the system to account.

Deliberation tells data holders what affected communities want. Leverage is what makes wanting matter—and what makes a social license a license at all, rather than a survey result. Extending the theory of social license in this direction does not diminish the value of participation; it completes it, by insisting that the invitation to be heard be matched by the capacity to be heeded.

## Bibliography

Addo, Peter, Adam Zable, Andrew J. Zahuranec, and Stefaan Verhulst. Reimagining Data Governance for AI. Technical Reports no. 78. Paris: Agence Française de Développement, 2025. https://www.afd.fr/en/ressources/reimagining-data-governance-ai.

Arnstein, Sherry R. “A Ladder of Citizen Participation.” Journal of the American Institute of Planners 35, no. 4 (1969): 216–24. https://doi.org/10.1080/01944366908977225.

Balkin, Jack M. “Information Fiduciaries and the First Amendment.” UC Davis Law Review 49, no. 4 (2016): 1183–1234.

Ben-Shahar, Omri, and Carl E. Schneider. More Than You Wanted to Know: The Failure of Mandated Disclosure. Princeton, NJ: Princeton University Press, 2014.

Birhane, Abeba, William Isaac, Vinodkumar Prabhakaran, Mark Díaz, Madeleine Clare Elish, Iason Gabriel, and Shakir Mohamed. “Power to the People? Opportunities and Challenges for Participatory AI.” In Proceedings of the 2nd ACM Conference on Equity and Access in Algorithms, Mechanisms, and Optimization (EAAMO '22), art. 6. New York: ACM, 2022. https://doi.org/10.1145/3551624.3555290.

Carroll, Stephanie Russo, Ibrahim Garba, Oscar L. Figueroa-Rodríguez, Jarita Holbrook, Raymond Lovett, Simeon Materechera, Mark Parsons, et al. “The CARE Principles for Indigenous Data Governance.” Data Science Journal 19, no. 1 (2020): art. 43. https://doi.org/10.5334/dsj-2020-043.

Centre for Information Policy Leadership. The Limitations of Consent as a Legal Basis for Data Processing in the Digital Society. CIPL, December 2024. https://www.informationpolicycentre.com/wp-content/uploads/2024/12/cipl_bkl_limitations_of_consent_legal_basis_data_processing_dec24-4.pdf.

Chamberlain, Neil W. Collective Bargaining. New York: McGraw-Hill, 1951.

Cybernews. “‘Data Strike’ Calls for Logging Off Social Media and Streaming Services on May 1st.” April 22, 2026. https://cybernews.com/tech/data-strike-may/.

Delacroix, Sylvie, and Neil D. Lawrence. “Bottom-Up Data Trusts: Disturbing the ‘One Size Fits All’ Approach to Data Governance.” International Data Privacy Law 9, no. 4 (2019): 236–52. https://doi.org/10.1093/idpl/ipz014.

Directive (EU) 2020/1828 of the European Parliament and of the Council of 25 November 2020 on representative actions for the protection of the collective interests of consumers and repealing Directive 2009/22/EC. OJ L 409, 4.12.2020, 1. https://commission.europa.eu/law/law-topic/consumer-protection-law/representative-actions-directive_en.

Dryzek, John S. Deliberative Democracy and Beyond: Liberals, Critics, Contestations. Oxford: Oxford University Press, 2000.

DTPR (Digital Trust for Places & Routines). Stewarded by Helpful Places. Accessed August 25, 2026. https://dtpr.io/.

Edwards, Lilian, and Michael Veale. “Slave to the Algorithm? Why a ‘Right to an Explanation’ Is Probably Not the Remedy You Are Looking For.” Duke Law & Technology Review 16 (2017): 18–84.

Felstiner, William L. F., Richard L. Abel, and Austin Sarat. “The Emergence and Transformation of Disputes: Naming, Blaming, Claiming…” Law & Society Review 15, no. 3/4 (1980–81): 631–54. https://doi.org/10.2307/3053505.

Frey, Seth, and Nathan Schneider. “Effective Voice: Beyond Exit and Affect in Online Communities.” arXiv preprint arXiv:2009.12470, September 25, 2020. https://arxiv.org/abs/2009.12470.

Fung, Archon. “Varieties of Participation in Complex Governance.” Public Administration Review 66, s1 (2006): 66–75. https://doi.org/10.1111/j.1540-6210.2006.00667.x.

Fung, Archon, and Erik Olin Wright. “Thinking about Empowered Participatory Governance.” In Deepening Democracy: Institutional Innovations in Empowered Participatory Governance, edited by Archon Fung and Erik Olin Wright, 3–42. London: Verso, 2003.

Global Indigenous Data Alliance. “CARE Principles for Indigenous Data Governance.” 2018. https://www.gida-global.org/careprinciples.

GovLab, The. “Updated and Expanded Selected Readings on Indigenous Data Sovereignty.” The GovLab Blog, October 12, 2020; updated October 11, 2021. https://blog.thegovlab.org/post/selected-readings-on-indigenous-data-sovereignty.

Hays, Paul R. Review of Collective Bargaining, by Neil W. Chamberlain. Indiana Law Journal 27, no. 2 (Winter 1952): art. 10. https://www.repository.law.indiana.edu/ilj/vol27/iss2/10.

Helpful Places. “Canada's AI strategy focuses on trust and transparency. But for whom?” June 30, 2026. https://www.helpfulplaces.com/insights/canada-s-ai-strategy-names-trust-and-transparency

Henderson, Jack, and Matt Prewitt. Data Coalitions & Escrow Agents. RadicalxChange Foundation, June 2023. https://www.radicalxchange.org/updates/documents/data-coalitions-and-escrow-agents.pdf.

Hirschman, Albert O. Exit, Voice, and Loyalty: Responses to Decline in Firms, Organizations, and States. Cambridge, MA: Harvard University Press, 1970.

International Network on Digital Self-Determination. Accessed August 25, 2026. https://www.idsd.network/.

Kaminski, Margot E. “Understanding Transparency in Algorithmic Accountability.” In The Cambridge Handbook of the Law of Algorithms, edited by Woodrow Barfield, 121–38. Cambridge: Cambridge University Press, 2020.

Khan, Lina M., and David E. Pozen. “A Skeptical View of Information Fiduciaries.” Harvard Law Review 133, no. 2 (2019): 497–541.

Levine, Peter. “Exit, Voice and Loyalty.” Civic Theory and Practice (blog), Tufts University, January 27, 2022. https://sites.tufts.edu/civicstudies/2022/01/27/exit-voice-and-loyalty/.

Mitchell, Margaret, Simone Wu, Andrew Zaldivar, Parker Barnes, Lucy Vasserman, Ben Hutchinson, Elena Spitzer, Inioluwa Deborah Raji, and Timnit Gebru. “Model Cards for Model Reporting.” In Proceedings of the Conference on Fairness, Accountability, and Transparency (FAT* '19), 220–29. New York: ACM, 2019. https://doi.org/10.1145/3287560.3287596.

Mulligan, Deirdre K., and Kenneth A. Bamberger. “Procurement as Policy: Administrative Process for Machine Learning.” Berkeley Technology Law Journal 34, no. 3 (2019): 773–852. https://doi.org/10.15779/Z38XG9FB0S.

New Commons Incubator for Indigenous Languages and Cultures. Accessed August 25, 2026. https://newcommons.ai/.

Open Data Policy Lab. "A Facilitator's Guide to Establishing a Social License for Data Reuse." February 10, 2026. https://opendatapolicylab.org/articles/new-publication-a-facilitators-guide-to-establishing-a-social-license-for-data-reuse/.

RadicalxChange Foundation. "Data Escrows." Accessed August 25, 2026. https://www.radicalxchange.org/projects/data-escrows/.

RadicalxChange Foundation. "The Data Freedom Act." Accessed August 25, 2026. https://www.radicalxchange.org/projects/the-data-freedom-act/.

Scott, James C. Seeing Like a State: How Certain Schemes to Improve the Human Condition Have Failed. New Haven, CT: Yale University Press, 1998.

Sieber, Renée, and Roberta Du. "Citizen Resistance to AI." Presentation at the Participatory AI Research & Practice Symposium (PAIRS) 2026, New Delhi and online, May 2026. https://aifortherestofus.ca/wp-content/uploads/2026/05/PAIRS-Sieber-2026.pdf.

Simpson, Audra. Mohawk Interruptus: Political Life across the Borders of Settler States. Durham, NC: Duke University Press, 2014.

Sloane, Mona, Emanuel Moss, Olaitan Awomolo, and Laura Forlano. "Participation Is Not a Design Fix for Machine Learning." In Proceedings of the 2nd ACM Conference on Equity and Access in Algorithms, Mechanisms, and Optimization (EAAMO '22), art. 1. New York: ACM, 2022. https://doi.org/10.1145/3551624.3555285.

Sterckx, Sigrid, Vojin Rakic, Julian Cockbain, and Pascal Borry. "'You Hoped We Would Sleep Walk into Accepting the Collection of Our Data': Controversies Surrounding the UK care.data Scheme and Their Wider Relevance for Biomedical Research." Medicine, Health Care and Philosophy 19, no. 2 (2016): 177–90. https://doi.org/10.1007/s11019-015-9661-6.

Tarkowski, Alek. NOODL: An Experiment in Equitable Data Licensing — Promise and Limits. Open Future, April 21, 2026. https://openfuture.eu/publication/noodl-an-experiment-in-equitable-data-licensing-promise-and-limits/.

Verhulst, Stefaan. "Data Stewardship Decoded: Mapping Its Diverse Manifestations and Emerging Relevance at a Time of AI." SSRN Scholarly Paper no. 5124555, February 2025. https://papers.ssrn.com/sol3/papers.cfm?abstract_id=5124555.

Verhulst, Stefaan. "Orchestrating and Designing Data Collaboratives: What Governance Model Is Fit for Purpose?" SSRN Scholarly Paper no. 6364718, March 7, 2026. https://papers.ssrn.com/sol3/papers.cfm?abstract_id=6364718.

Verhulst, Stefaan, and Adam Zable. Operationalizing a Social License for Data Re-Use: A Facilitator's Guide and Worksheet — Questions to Signal and Capture Community Preferences and Expectations. New York: The GovLab / Open Data Policy Lab, 2026. https://tools.opendatapolicylab.org/files/Data-Re-Use-Worksheet.pdf.

Verhulst, Stefaan, Andrew J. Zahuranec, Hannah Chafetz, Leona Verdadero, and Jennifer Hansen. Data Commons: Frequently Asked Questions. New Commons Incubator, in collaboration with the Open Data Policy Lab at The GovLab, 2026. https://opendatapolicylab.org/articles/data-commons-faq-report/.

Viljoen, Salomé. “A Relational Theory of Data Governance.” Yale Law Journal 131, no. 2 (2021): 573–654. https://papers.ssrn.com/sol3/papers.cfm?abstract_id=3727562.

Vincent, Nicholas, and Brent Hecht. “Can ‘Conscious Data Contribution’ Help Users to Exert ‘Data Leverage’ against Technology Companies?” Proceedings of the ACM on Human-Computer Interaction 5, no. CSCW1 (April 2021): art. 103. https://doi.org/10.1145/3449177.

Vincent, Nicholas, Brent Hecht, and Shilad Sen. “‘Data Strikes’: Evaluating the Effectiveness of a New Form of Collective Action against Technology Companies.” In The World Wide Web Conference (WWW '19). New York: ACM, 2019. https://doi.org/10.1145/3308558.3313742.

Vincent, Nicholas, Hanlin Li, Nicole Tilly, Stevie Chancellor, and Brent Hecht. “Data Leverage: A Framework for Empowering the Public in Its Relationship with Technology Companies.” In Proceedings of the 2021 ACM Conference on Fairness, Accountability, and Transparency (FAccT '21), 215–27. New York: ACM, 2021. https://doi.org/10.1145/3442188.3445885.

Winter, Jenifer Sunrise, and Elizabeth Davidson. “Governance of Artificial Intelligence and Personal Health Information.” Digital Policy, Regulation and Governance 21, no. 3 (2019): 280–90. https://doi.org/10.1108/DPRG-08-2018-0048.

Zong, Jonathan, and J. Nathan Matias. “Data Refusal from Below: A Framework for Understanding, Evaluating, and Envisioning Refusal as Design.” ACM Journal on Responsible Computing 1, no. 1 (2024): art. 10. https://doi.org/10.1145/3630107.